\documentclass[aps,prl, twocolumn,groupedaddress,showpacs]{revtex4-1}

\usepackage{graphicx}% Include figure files
\usepackage{color}
\usepackage{dcolumn}% Align table columns on decimal point
\usepackage{bm}% bold math
\begin{document}

% Use the \preprint command to place your local institutional report
% number in the upper righthand corner of the title page in preprint mode.
% Multiple \preprint commands are allowed.
% Use the 'preprintnumbers' class option to override journal defaults
% to display numbers if necessary
%\preprint{}

%Title of paper
\title{Bimodal island size distribution in heteroepitaxial growth}

% repeat the \author .. \affiliation  etc. as needed
% \email, \thanks, \homepage, \altaffiliation all apply to the current
% author. Explanatory text should go in the []'s, actual e-mail
% address or url should go in the {}'s for \email and \homepage.
% Please use the appropriate macro foreach each type of information

% \affiliation command applies to all authors since the last
% \affiliation command. The \affiliation command should follow the
% other information
% \affiliation can be followed by \email, \homepage, \thanks as well.
\author{P. V. Chinta}
\author{R. L. Headrick}
%\email[]{Your e-mail address}
%\homepage[]{Your web page}
%\thanks{}
%\altaffiliation{}
\affiliation{Department of Physics, University of Vermont, Burlington, VT 05405}

%Collaboration name if desired (requires use of superscriptaddress
%option in \documentclass). \noaffiliation is required (may also be
%used with the \author command).
%\collaboration can be followed by \email, \homepage, \thanks as well.
%\collaboration{}
%\noaffiliation

\date{\today}

\begin{abstract}
% insert abstract here
A bimodal size distribution of two dimensional islands is inferred during interface formation in heteroepitaxial growth of Bismuth Ferrite on (001) oriented SrTiO$_3$ by sputter deposition. Features observed by in-situ x-ray scattering  are explained by a model where coalescence of islands determines the growth kinetics with negligible surface diffusion on SrTiO$_3$.  Small clusters maintain a compact shape as they coalesce, while clusters beyond a critical size impinge to form large irregular connected islands and a population of smaller clusters forms in the spaces between the larger ones.    

% The second and subsequent layers nucleate first on the large islands, resulting in replication of the large island structure as layer-by-layer growth proceeds.
\end{abstract}

% insert suggested PACS numbers in braces on next line
\pacs{61.05.cf, 68.47.Gh, 68.55.A-, 81.10.Aj, 81.15.Cd, 81.15.Kk}
%\maketitle must follow title, authors, abstract, \pacs, and \keywords
\maketitle

% body of paper here - Use proper section commands
% References should be done using the \cite, \ref, and \label commands

%\section{\label{sec:Intro}Introduction}

%Check to see if the reference to Section \ref{sec:Intro} works.
Control of atomic-level processes in heteroepitaxial thin film growth is critically  important for the formation of interfaces in artificially layered nanoscale structures.  In turn, growth modes determine or influence important interface properties such as roughness, chemical intermixing, defects, and strain.  Phenomena typically observed in {\em{homoepitaxy}} arise from well-known processes of random atomic deposition, surface diffusion,  and the aggregation and coalescence of two-dimensional (2D) clusters.\cite{AMAR:1995fk}  At moderate growth temperatures, these processes lead to layer-by-layer (LBL) crystal growth. \cite{Blank:1999ys, Ferguson:2009fk}  In {\em{heteroepitaxy}}, defined as layered crystal growth of two or more materials with compatible crystal structures and lattice constants, there are other modes than can be observed.\cite{Opel:2012mz,Chambers:2000fr} The best known of these are three-dimensional (Volmer-Weber) and 2D followed by a transition to three-dimensional (Stranski-Krastanov). However, there may be additional possible modes involving only 2D structures during LBL growth. 

Here, we discuss a case of heteroepitaxy where interface formation is dominated by coalescence of 2D clusters.   In the case we will consider, surface diffusion on the substrate surface is very low so that  the mobility of single monomers can be neglected.   However, surface diffusion of deposited monomers that land on the overlayer and at the boundaries of overlayer islands is fast in comparison.  This leads to efficient coalescence of compact 2D islands over a range of length scales, and the system exhibits kinetics that are more akin to droplet growth processes (Family and Meakin, Blackman and Brochard, Refs.~\onlinecite{Family:1988kx,Family:1989fj,Blackman2000}) than to standard surface diffusion driven aggregation and coalescence.

 \begin{figure}
 \includegraphics[width=3.25 in]{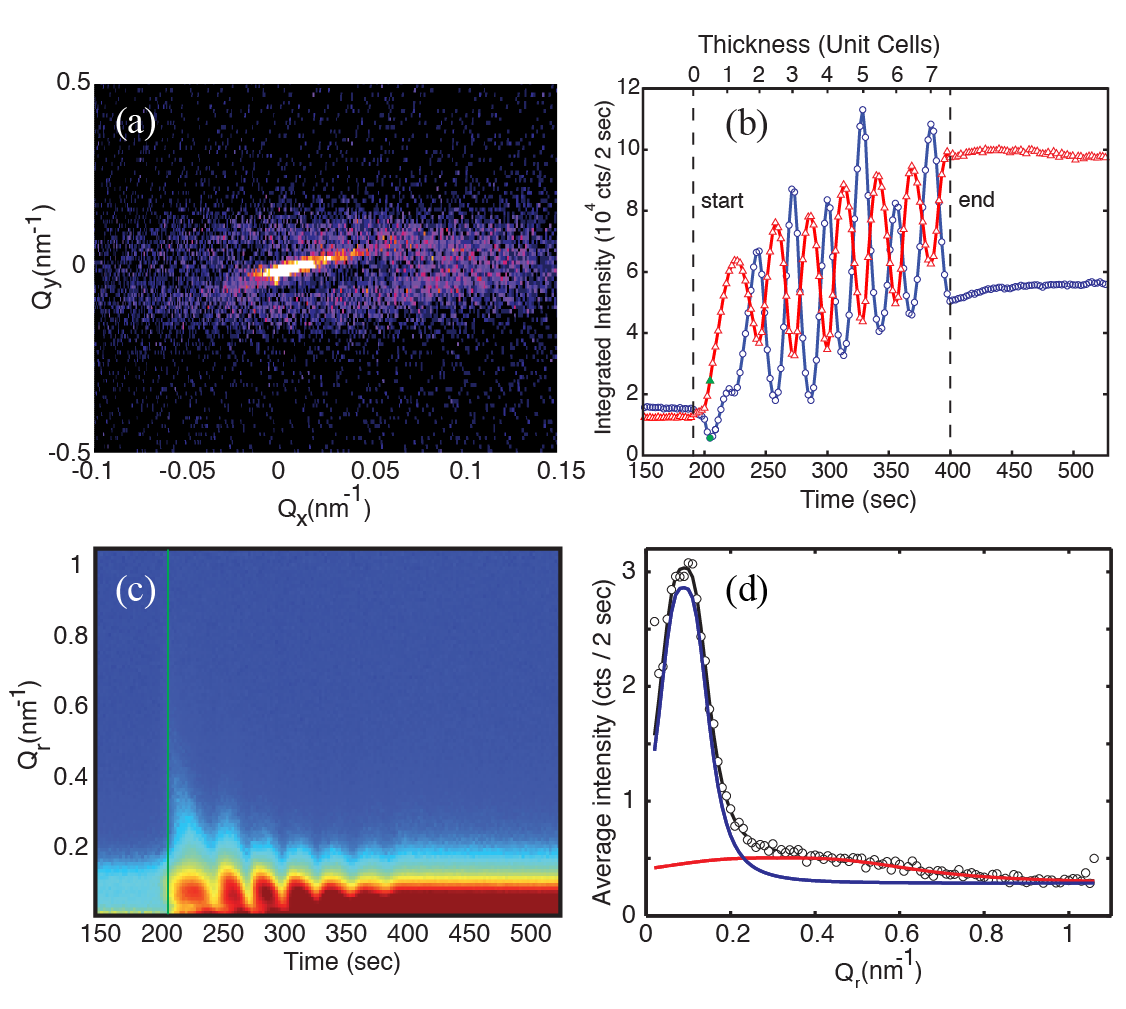}
 \caption{\label{fig:Fig1} X-ray reflectivity data near the (0 0 0.5) reflection during RF sputter deposition of 7.5 UC of BFO. The growth temperature and pressure were  650$^\circ$C and 20 mTorr, respectively  (with Ar:O$_2$  of 2:1). (a) Two dimensional image at a nominal film thickness of 0.5 UC.   (b) Specular (blue circles) and diffuse (red circles) integrated intensities.  (c) Time resolved diffuse scattering map of circularly averaged  profiles versus time.  (d) Circularly averaged intensity at $\theta = 0.5$.  The data (open circles) is fit  using a two component empirical expression (lines).  Data points at 205 s in (b) have been filled in green, and a corresponding green line in (c) mark the time slice corresponding to panels (a) and (d). 
}
 \end{figure}

In the case described above, clusters should theoretically grow exponentially with deposition time.\cite{Family:1988kx} The process would  lead to a single cluster covering the entire sample surface, except for kinetic limitations that set in at a material dependent yet well-defined length scale. Growth models incorporating the effects of kinetically limited coalescence have previously been developed. For example, the Interrupted Coalescence Model (ICM) and the Kinetic Freezing Model (KFM) successfully reproduce irregular or fractal patterns observed in vapor deposited metal thin films on inert substrates.\cite{Yu1991,Jeffers1994, Voss:1982uq}   Similar observations have been reported for ultra-thin epitaxial metal films that initially grow in a Volmer-Weber mode on single-crystal oxide substrates, followed by coalescence into islands with a distinct bimodal distribution.\cite{Zhang:1999qf}

In this letter we show that the ICM can be adapted to describe experimental observations of the layer-by-layer growth process in a case where the predictions of standard LBL growth models fail.    One important prediction of ICM is a bimodal distribution of  2D cluster sizes in good agreement with the experimental data.  This model may find wide applicability in cases where there is a  disparity in surface diffusion coefficients between the substrate and the film. An intriguing example is for  SrRuO$_3$ growth on SrTiO$_3$  where a change of the surface termination during the first growth layer leads to a large enhancement of surface diffusion of monomers on the overlayer.\cite{Rijnders2004}

% \section{\label{sec:Experimental}Experimental}

BFO  has attracted much interest due to its high ferroelectric polarization, coupled with antiferromagnetism and weak ferromagnetism.\cite{Wang:2003fk, Catalan2009}.   For this study, epitaxial  films were grown on  TiO$_2$-terminated (001) SrTiO$_3$ (STO) substrates using on-axis radio-frequency magnetron sputter deposition in a custom growth chamber situated at beamline X21 at the National Synchrotron Light Source. Film growth was monitored by in-situ x-ray scattering using radiation with $\lambda = 0.124$ nm. A fast single-photon counting x-ray area detector was used to simultaneously record the evolution of specular and diffuse intensities near the anti-Bragg scattering condition.   The morphology of the final surfaces were also corroborated with ex situ atomic force microscopy (AFM) measurements and by additional x-ray diffraction measurements.

% \section{\label{sec:Results}Results}

 \begin{figure}
 \includegraphics[width=2.6 in]{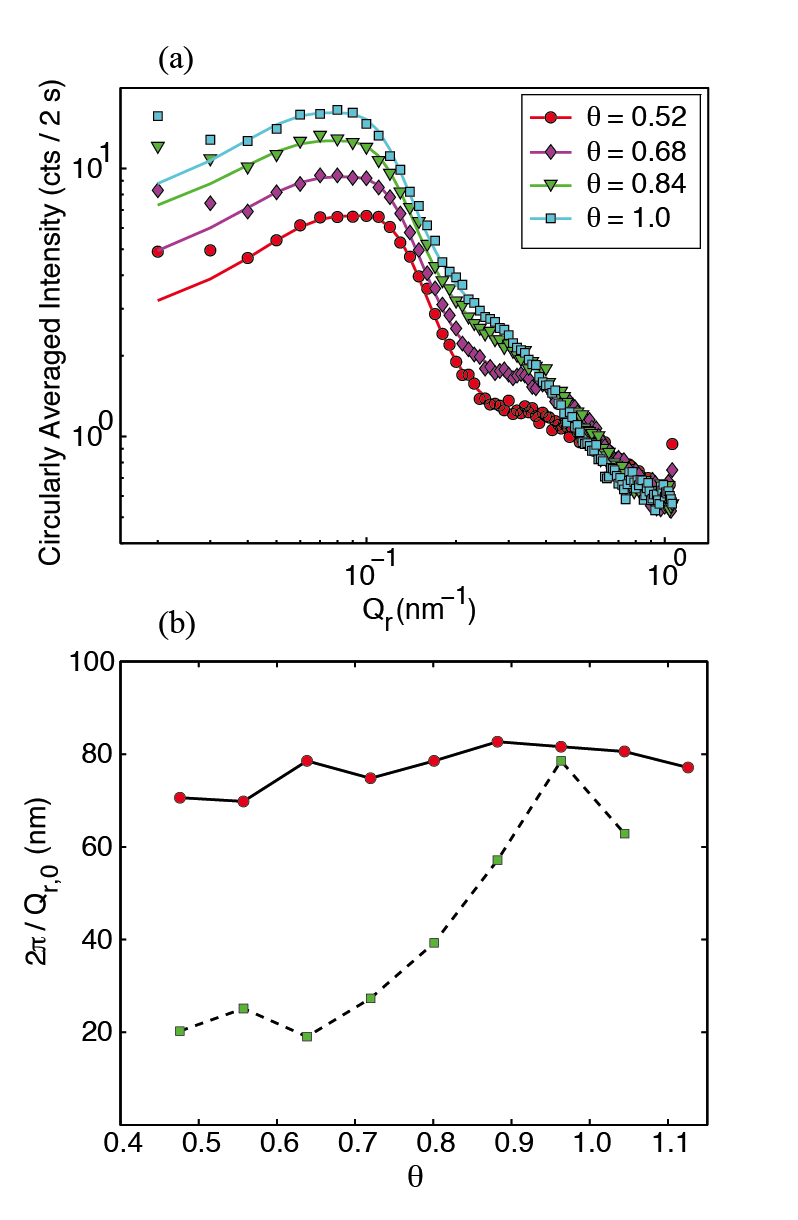}
 \caption{\label{fig:Fig2}(a) Diffuse scattering line shapes (circles) during coalescence of the first layer $0.5 \le \theta \le 1.0$ for BFO growth. The solid lines are two component fits to the data. (b) Estimated island separation obtained from the peak positions of the diffuse lobes. Low $Q_r$ component:  red circles and solid line;  high $Q_r$ component: green circles and dashed line.
}
 \end{figure}

 \begin{figure*}
 \includegraphics[width=6.0 in]{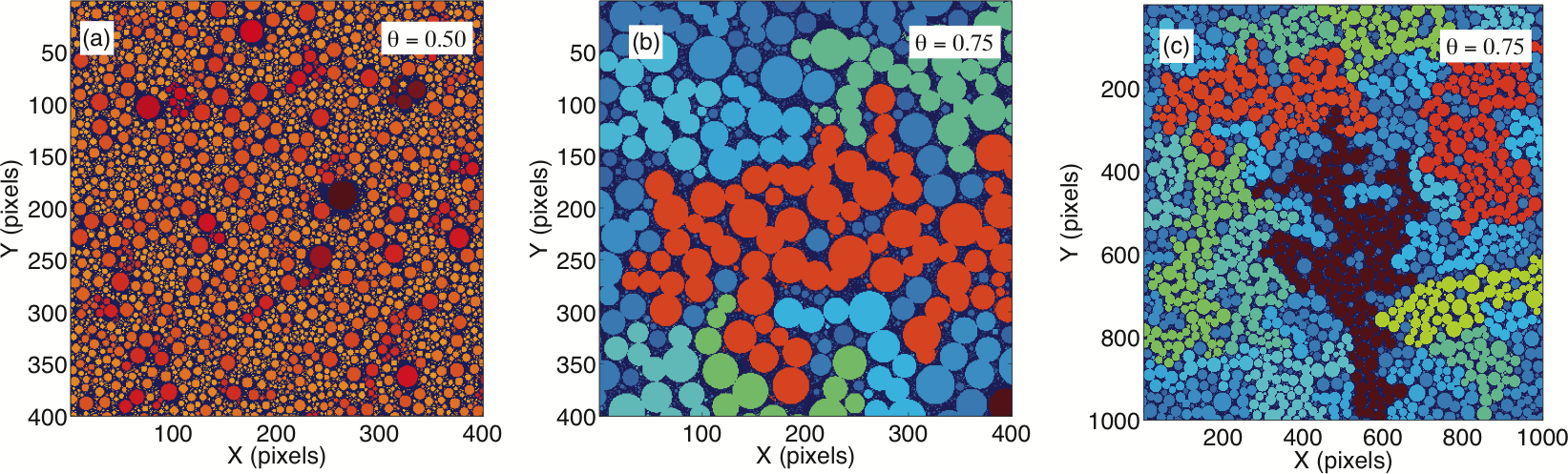}
 \caption{\label{fig:Fig3}Simulated cluster maps for ICM on a $1000 \times 1000$ grid with $D=2$ and $\pi R_c^2 = 200$ monomers. (a) shows that at 0.50, the clusters are still mainly isolated. (b) and (c) show large connected regions just before the percolation coverage is reached.   The color scale in each image  is by the size of connected regions. }
 \end{figure*}

Fig.~\ref{fig:Fig1}(a) shows a time slice from a series of images recorded during BFO deposition and corresponds to 0.5 unit cell (UC) nominal film thickness.  The image shows an almost perfectly circular diffuse ring that forms  around the specular reflection, indicating the presence of correlated 2D islands on the surface.   The specular spot near the center of the image is elongated due to the terrace structure of the STO substrate, where a step spacing of $\approx700$ nm was observed by AFM. Fig.~\ref{fig:Fig1}(b) shows the integrated specular and diffuse intensities at (0 0 0.5) for deposition to 7.5 UC of BFO.  In this letter, we focus on the the submonolayer deposition regime.  An important feature to note is that while the specular intensity reaches a minimum at coverage $\theta = $ 0.5, the diffuse intensity continues to increase monotonically up to $\theta = $ 1.5.   This unusual behavior is due to the late peaking of the diffuse scattering during growth of the first unit cell, which we will explain in detail below, and also due to the nucleation of the second layer before the completion of the first layer.

More detail is obtained in Fig.~\ref{fig:Fig1}(c), where circularly averaged $Q_r$ radial profiles vs. time are presented. The {\em{absence}} of any   strong features below $\theta = 0.5$ indicates that the nuclei formed on the surface are very small, $<5$ nm.   Broad features extending up to $Q_r \approx 0.5$ nm$^{-1}$ are found to appear  for $\theta \ge 0.5$,  indicating formation or coarsening of clusters at very short length scales $\le 20$ nm.   The strong ``ring" feature also becomes visible at about the same coverage.    The evolution of the diffuse intensity  for $\theta > 1$ appears to be compatible with the standard LBL growth mode via surface diffusion and aggregation. However, our observations for $\theta < 1$ are inconsistent with standard LBL since it predicts strong diffuse scattering in the aggregation regime, {\em{peaking}} in intensity at $\approx 0.5$.

Fig.~\ref{fig:Fig1}(d) shows the radial profile for a single frame at $\theta = 0.5$. The data is fit using the empirical form suggested by Brock et al.\cite{Brock2010}

\begin{equation}
I(Q_r) = \frac{I_0}{\left[  1 + \xi^2 (Q_r - Q_{r,0})^2   \right]^{3/2}} \label{eq:Eq1}
\end{equation}
where $Q_{r,0}$ determines the peak position, and $\xi$ the peak width. Two different diffuse components are observed at $Q_{r,01} = 0.09$ nm$^{-1}$ and $Q_{r,02} = 0.31$ nm$^{-1}$ indicating surface features of different length scales. The separation between the two components is considerably larger than what we expect for a single population of disk-like features on the surface, based on calculated structure factors. Specifically, the disk structure factor $S(Q_r) \propto [J_0(RQ_r)/RQ_r]^2$ produces a series of fringes, but these are too closely spaced to produce both components observed in the data of Fig.~\ref{fig:Fig1}. In this expression, $R$ is the disk radius and $J_0$ is the Bessel function of the first kind. Fig.~\ref{fig:Fig2} shows how the peak positions and the estimated length scales evolve with coverage between 0.5 and 1.0 UC. The results show that the length scales are relatively constant for $\theta \le 0.65$, and, only the broad component coarsens for $\theta > 0.65$. The lack of significant coarsening of the sharp component  indicates that some mechanism plays a role to prevent the coarsening of large islands. Below, we discuss these results in terms of a coalescence-dominated model.

 %\section{\label{sec:SimResults}Simulation Results}

Standard models of  LBL growth generally involves three regimes: nucleation ($\theta \le 0.1$), aggregation ($0.1 < \theta < 0.4$), and coalescence ($\theta > 0.5$).\cite{AMAR:1994kx} Initially, deposited monomers diffuse on the substrate, and a stable nucleus is formed when a critical number of them meet. Once a high enough density of nuclei is reached, the monomer density drops dramatically and the nucleation rate drops correspondingly.  Aggregation thus refers to the growth of existing clusters at a nearly fixed number density. Finally in the coalescence regime, the islands begin to join together and eventually  form  a continuous layer.  Impingement is  a special case of coalescence where the redistribution of matter among islands does not take place after their collision.\cite{Tomellini2006}

 Our experimental observations lead us to a different model: (a) very little  surface diffusion, producing small length scales in the early stages of monolayer formation; (b) formation of compact clusters on the surface, so that the asymptotic form of the structure factor is $S(Q_r)\propto Q_r^{-3}$ as $Q_r\rightarrow \infty$, as for disks; (c) irreversible attachment of monomers to the islands, since relaxation effects after deposition is stopped are minimal. In order to model this process, we have performed Monte-Carlo simulations on a $1000 \times 1000$ array.    Clusters are assumed to be perfectly compact disks with irreversible monomer attachment, and monomers landing atop existing islands migrate instantaneously to the island edge. We assume the critical cluster size $i = 0$ case with no diffusion.  Consequently, there is no aggregation regime and coalescence effects dominate for all coverages. 
 
The FM model has been studied for surfaces of dimension $d$ and droplets of dimension $D$ for many combinations with $D\ge d$.\cite{Family:1988kx,Family:1989fj,Blackman2000}  Our experiments relate to the case D = d = 2, i.e. two dimensional clusters on a 2D surface. This leads to a situation where the mean cluster size grows exponentially.  This behavior is inconsistent with our BFO radial profiles, where the low-Q peak shifts very little with coverage.  In addition, FM with $D=2$ does not lead to a bimodal distribution of cluster sizes and we find that the structure factors produced by this model do not have a pronounced sharp component as we have observed in our experiment [Fig.~\ref{fig:Fig1}(d)].

 \begin{figure*}
 \includegraphics[width=6.0 in]{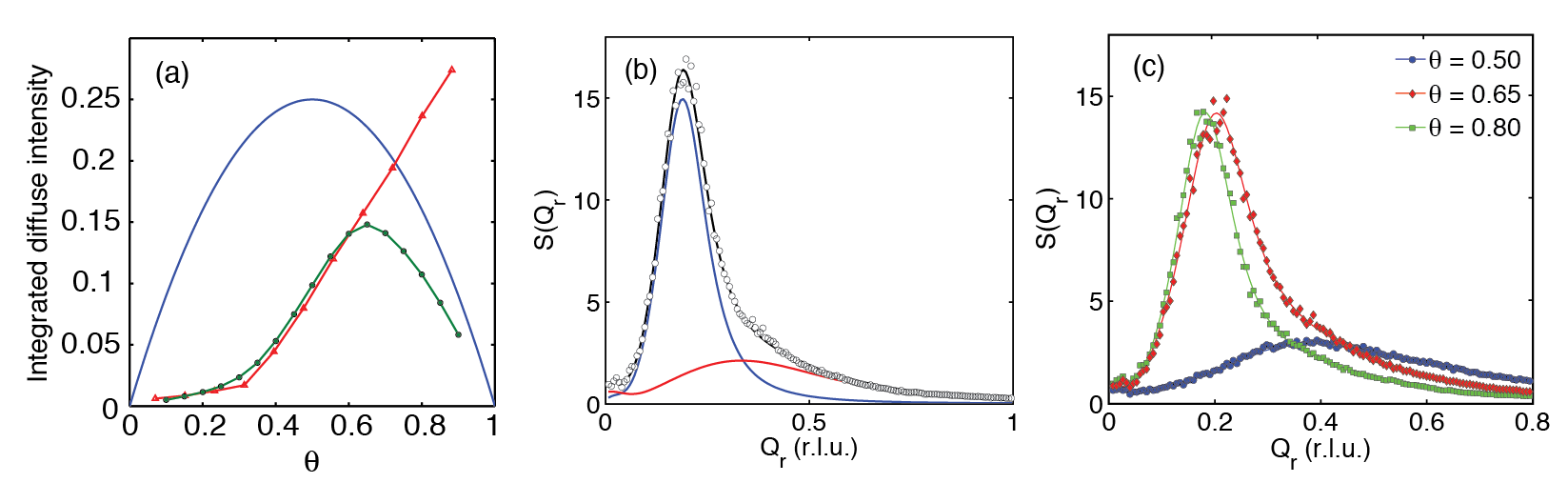}
 \caption{\label{fig:Fig4}Calculated structure factors for ICM. (a) illustrates that rather than peaking at $\theta=0.5$ as the total diffuse scattering intensity does (line), the integrated intensity with $Q_r < 0.8$ reciprocal lattice units (r.l.u.) exhibits a delayed peak at $\theta=0.65$ (circles).  For comparison, the triangles show the rescaled total diffuse intensity from Fig. 1(b). Plot (b) shows  the two component fit at $\theta=0.70$.    (c) shows the evolution of the lineshape at three coverages, illustrating the sudden appearance of the narrow component for coverages above $\theta = 0.5$, and  sharpening of the line shape near $\theta = 0.8$.}
 \end{figure*}

After considering several possible mechanisms to limit the growth of the largest islands, we decided to add the ICM mechanism to our model.\cite{Yu1991}  Monomers (or particles with a size $R_0$) are added to the surface in random locations, as in FM. However, once clusters reach a certain critical size $R_c$ they no longer coalesce by merging with each other, rather the clusters impinge without combining.  Deposited monomers and small clusters below the cutoff are still allowed to combine with larger clusters.  This  model results in the formation of irregular connected islands composed of impinging 2D clusters.   

Fig.~\ref{fig:Fig3}(a) shows ICM results for $\theta = 0.5$.   At this stage, the great majority of clusters have not reached the cutoff size.  Small clusters are continually replenished because when clusters merge their centers move together, exposing a region of the surface for new clusters to nucleate.  This regeneration effect is central to the FM mechanism.   Fig.~\ref{fig:Fig3}(b,c) show two views of a cluster map for $\theta = 0.75$. At this stage the largest clusters have formed connected islands, while a second population of smaller clusters continues to develop within the interstices of the larger ones. Thus, a feature of ICM with $D=2$ is the formation of a bimodal cluster distribution.  It is caused by a depletion of clusters just below the cutoff size, which are most likely to collide and coalesce with larger clusters.   We also observe that  $\theta = 0.75$ is close to the percolation threshold $\theta_p$, since the largest connected region nearly spans the map. This is in agreement with the results of Yu et al., who find $\theta_p \approx 0.78$ for ICM with $R_c/R_0 = 4$.\cite{Yu1991}

Fig.~\ref{fig:Fig4} shows results for structure factors generated from ICM cluster maps. Fig.~\ref{fig:Fig4}(a) shows the total diffuse scattering as a function of coverage up to $\theta = 1$, as well as the integrated intensity within a region of $Q_r$ meant to illustrate the diffuse intensity striking a detector of limited size.  At the early stages when clusters are very small, a small fraction of the total intensity reaches the detector.  This behavior reproduces our experiment, where very little diffuse signal is detected for $\theta < 0.5$, and the peak occurs late so that it merges into the diffuse signal after the second layer has nucleated.  We have included the experimental diffuse data from Fig 1(b) in Fig. 4(a) for comparison, which is consistent with continued coarsening for $\theta > 0.65$ as shown in Fig. 2, implying that the approximation of instantaneous island coalescence is too drastic. Fig.~\ref{fig:Fig4}(b) shows a two component fit of the radial profile of the structure factor for $\theta = 0.7$, where a pronounced second component is observed.  Fig.~\ref{fig:Fig4}(c) shows the evolution of the lineshape for disconnected islands ($\theta = 0.5$), connected islands with a strong component from the smaller  islands ($\theta = 0.65$), and at percolation where a significant fraction of the small islands have merged with the connected regions ($\theta = 0.8$).  The results reproduce the sudden appearance of the sharp peak, which has been one of the most puzzling aspects of our experimental data. We find that tuning $R_c$ has little effect on the shape of $S(Q_r)$ at a given coverage, but simply changes the overall length scale.  

To conclude, we find that a coalescence-dominated model explains the structural evolution during interface formation in BFO layer-by-layer growth on STO(001).  The growth mode is distinguished from standard layer-by-layer growth by a bimodal cluster size distribution, which we have observed experimentally and confirmed through simulations.  

\begin{acknowledgments}
This work was supported by the U.S. DOE Office of Science, Office of Basic Energy Sciences under DE-FG02-07ER46380. Use of the NSLS was supported by the U.S. DOE, Office of Science, Office of Basic Energy Sciences, under Contract No. DE-AC02-98CH10886.
\end{acknowledgments}

% Create the reference section using BibTeX:
%\bibliographystyle{Unsrt}
%\bibliography{Priya_BFO_Sputter_V2}

%

\end{document}